# Hyperspectral Image Classification Method Based on Spiking Neural Network

Yang Liu*, Yahui Li, Rui Li, Liming Zhou, Lanxue Dang, Huiyu Mu, and Qiang Ge*

*Abstract*—Convolutional neural network (CNN) performs well in Hyperspectral Image (HSI) classification tasks, but its high energy consumption and complex network structure make it difficult to directly apply it to edge computing devices. At present, spiking neural networks (SNN) have developed rapidly in HSI classification tasks due to their low energy consumption and event driven characteristics. However, it usually requires a longer time step to achieve optimal accuracy. In response to the above problems, this paper builds a spiking neural network (SNN-SWMR) based on the leaky integrate-and-fire (LIF) neuron model for HSI classification tasks. The network uses the spiking width mixed residual (SWMR) module as the basic unit to perform feature extraction operations. The spiking width mixed residual module is composed of spiking mixed convolution (SMC), which can effectively extract spatial-spectral features. Secondly, this paper designs a simple and efficient arcsine approximate derivative (AAD), which solves the non-differentiable problem of spike firing by fitting the Dirac function. Through AAD, we can directly train supervised spike neural networks. Finally, this paper conducts comparative experiments with multiple advanced HSI classification algorithms based on spiking neural networks on six public hyperspectral data sets. Experimental results show that the AAD function has strong robustness and a good fitting effect. Meanwhile, compared with other algorithms, SNN-SWMR requires a time step reduction of about 84%, training time, and testing time reduction of about 63% and 70% at the same accuracy. This study solves the key problem of SNN based HSI classification algorithms, which has important practical significance for promoting the practical application of HSI classification algorithms in edge devices such as spaceborne and airborne devices.

## I. INTRODUCTION

In recent years, hyperspectral technology has developed rapidly. It has hundreds of continuous three-dimensional data of hyperspectral bands in the same scene and contains rich spectral information and spatial information. Therefore, HSI is widely used in geophysical exploration[1], urban planning[2], soil salinity estimation[3], environmental monitoring and forestry management[4], [5], [6], and other fields.

*A Traditional HSI classification methods.*

HSI classification aims to assign a unique semantic label to each pixel in HSI, which is the first and most important step in realizing hyperspectral remote sensing applications. The classification effect will directly affect the accuracy of information obtained from HSI, and then affect the subsequent engineering application results. Early HSI classification methods mainly include support vector machine (SVM)[7], multivariate logistic regression[8], dynamic or logical subspace[9], [10], neural networks[11], etc. At the same time, since the Hughes phenomenon[12] in HSI images greatly limits the ability of HSI classification algorithms, scholars have proposed a series of dimensionality reduction techniques, trying to reduce the data dimension through feature selection and feature extraction, such as linear discriminant analysis (LDA)[13], principal component analysis (PCA)[14], [15], independent component analysis (ICA)[16], [17], etc. Due to the problems of "homologous heterospectral" (that is, two identical objects may have different spectral features) and "foreign object homospectral" (that is, two different objects may have the same spectral features) in HSI, relying solely on spectral information cannot achieve good classification results. Therefore, using multi-view learning (MVL)[18] to combine different information and distinguish spectral features through spatial information such as shape and texture can effectively improve the accuracy of HSI classification. Subsequently, scholars proposed a series of classification methods based on spatial spectral information, such as extended morphological contours[19], [20], and joint sparse representation models[21]. However, the above methods usually use experts to design manual features for specific scenarios, which results in poor scene adaptability. Therefore, how to extract features automatically, accurately, and quickly is still the key to improving the HSI classification effect.

*B HSI classification methods based on CNN.*

To solve the above problems, deep learning technology has been applied by researchers to HSI classification tasks. Different from traditional algorithms, deep learning-based algorithms automatically extract high-level abstract features in HSI in a hierarchical learning manner through convolutional neural networks (CNN). In addition, its deep nonlinear network has more powerful feature representation capabilities, allowing it to achieve higher classification accuracy.

This work is supported by National Natural Science Foundation of China (No.62176087).
Yang Liu (Email: ly.sci.art@gmail.com), Henan Key Laboratory of Big Data Analysis and Processing, School of Computer and Information Engineering, Henan University, Kaifeng 475004, PR China. Yahui Li, Rui Li, Liming Zhou, Lanxue Dang, Huiyu Mu, and Qiang Ge are Henan Key Laboratory of Big Data Analysis and Processing, School of Computer and Information Engineering, Henan University, Kaifeng 475004, PR China.
(Corresponding authors: Qiang Ge and Yang Liu)

Early deep learning methods typically relied on stacked autoencoder (SAE)[22] and deep belief network (DBN) [23]. These methods were able to extract deep features, but they could disrupt spatial context correlation and cause the problem of spatial information loss. Therefore, researchers have proposed a series of CNN models[24], [25] to process three-dimensional image patches to extract more spatial context information and improve classification accuracy. Although CNN-based methods have achieved remarkable results in HSI classification tasks, the training of CNN requires a large amount of data, and there is less labeled data available in HSI. Generally, deeper networks can extract more information from less data, however, deep networks are often accompanied by the risk of overfitting [26]. In response to the above issues, He et al. [27] proposed a residual network (ResNet), which transfers parameters to deep layers through residual connections, which alleviates the over-fitting problem caused by deeper networks.

With the emergence of ResNet, researchers have built many deep networks for HSI tasks[28], [29], [30], [31]. Gao et al. [28] proposed a multi-scale residual network for HSI classification, which introduces mixed convolution in the residual network to extract features of different scales from multiple feature maps and aggregates shallow and deep features through residual connections. Li et al. [29] proposed a dual-channel CNN based on automatic clustering. They first reduced the variance between categories in the spectral dimension through automatic clustering and then used CNN to extract spectral dimension information. Zhong et al. [30] proposed a spatial residual network for HSI classification, which extracts spectral features and spatial features by constructing spatial residual blocks and spectral residual blocks. Wang et al. [31] designed a densely connected network that automatically extracts rich spatial and spectral features in HSI by constructing dense spectral blocks and dense spatial blocks, effectively improving HSI classification accuracy. The above methods have achieved excellent results in HSI classification, but their model design is complex and the network is deep. Generally, the depth of the network causes problems such as longer training and testing times and more parameters. At the same time, CNN operations require a large number of multiplication operations to complete, which will lead to increased computing costs and energy consumption. For the above reasons, current HSI classification algorithms are rarely used in real life. Therefore, a high-precision, low-latency network is of great significance for the practical application of HSI classification tasks.

*C HSI classification methods based on SNN.*

SNN is a new generation of the artificial neural network, which is similar to the human brain and has long-term development prospects due to its low power consumption, event-driven characteristics, and biological rationality. At the same time, unlike the CNN, SNN only uses addition operations and avoids multiplication[32]. Therefore, this paper aims to construct a lightweight, fast, and high-precision spiking neural network for HSI classification tasks to promote the application of HSI classification algorithms in unmanned autonomous equipment such as satellites and drones.

SNN can be divided into supervised learning and unsupervised learning. At present, unsupervised SNN are mainly trained through the STDP mechanism. STDP is a biologically interpretable local learning algorithm that determines the direction and magnitude of synaptic weight changes based on the relative time difference between pre-synaptic and post-synaptic peaks to update the weight to complete the training of SNN. However, due to limitations in computing resources, this algorithm is still limited to simple models[33], [34], [35].

There are two main categories of supervised SNN training methods: ANN to SNN and direct training methods. The idea of the ANN to SNN[36], [37] method is to obtain an SNN that can produce the same input and output mapping as a CNN under a given task[38], [39]. Currently, ANN to SNN is an effective solution to implement deep SNN. However, the inference time converted from ANN to SNN is very large, resulting in a decrease in energy efficiency and an increase in latency. Another method directly trains SNN. However, the update of weights in artificial neural networks is mainly based on the backpropagation algorithm. Since the spiking signal of SNN is discrete, the backpropagation algorithm cannot be directly used for network optimization.

In response to the problem that supervised SNN are difficult to train directly, researchers have tried to use approximate derivative algorithms to proxy gradients to achieve the purpose of optimizing the network[40], [41], [42], [43]. Wu et al. [40] and others used the mean square error function to derive the formulas of the time dimension and the space dimension, and for the first time, they transformed the LIF model and used the rectangular function approximate differential to train the SNN[41]. F Zenke et al. [42] system provides an overview of the concepts of synaptic plasticity and data-driven learning, investigates the influence of proxy gradient parameters on classification accuracy, analyzes the robustness of proxy derivatives with different shapes, and provides guidance for the research of SNN. Cheng et al. [43] combined membrane potential update and lateral effects of neurons to update local weights, improving the performance of spiking neural networks. The above methods are mainly aimed at natural images and are difficult to directly apply to HSI classification tasks.

Currently, to realize the application of HSI in unmanned autonomous equipment, researchers have proposed some HSI classification algorithms based on SNN[44], [45]. Liu et al. [44] designed approximate derivatives for the IF neuron model and built a SNN-SSEM network through the SE module for HSI classification. The algorithm achieved advanced accuracy at 100-time steps. Liu et al. [45] constructed a Square Approximate Derivative (SAD) for backpropagation for the LIF model and built a dual-branch structure SNN. This algorithm only requires 40-time or 50-time steps to achieve optimal accuracy. The above algorithms have achieved excellent results under long time steps. However, the length of the time step is positively correlated with the training time and testing time of the network. Therefore, constructing a high-precision, short-time-step spiking neural network is crucial for HSI classification tasks. In response to the above problems, this paper proposes a directly trained SNN for HSI classification tasks at short time steps. Our contributions are summarized as follows:

1) To solve the problem that supervised SNN is difficult to train directly, an arcsine approximate derivative (AAD) is proposed. This function realizes the backpropagation of the spiking neural network by fitting the Dirac function and can realize the direct training of the supervised spiking neural network through this function.
2) A spiking neural network (SNN-SWMR) that can be trained quickly and with high accuracy in a short time step is constructed for HSI classification tasks. The network can achieve optimal accuracy in a minimum of 10-time steps. To the best of our knowledge, this is the shortest required time step among HSI classification algorithms based on SNN.
3) We conducted experimental evaluations on six publicly available HSI datasets. The results show that compared with the most advanced HSI classification algorithms based on SNN, this network has significant advantages in terms of time step size, training time, and testing time.

## II. APPROACH

In this section, we will give a detailed introduction to the SNN-SWMR constructed in this paper for HSI classification and the proposed AAD for the backpropagation of SNN.

*A LIF Neuron*

The human brain is a complex and efficient network composed of billions of neurons, which can perform outstanding functions such as recognition and reasoning in just 20 watts[46]. Neurons are the fundamental units of brain computing[47], which transmit or exchange information through discrete action potentials or "spiking". Neurons process input information and convert the spiking into membrane potentials. The spiking neuron model converts neurons into RC circuits through resistance R and capacitance C[48], and its structure is shown in Fig. 1. When a spiking neuron receives an input current I, the charge is charged to the cell membrane, which acts as a capacitor C. Since the cell membrane cannot be regarded as an insulator, the cell membrane is regarded as a resistor R. The current will seep out of the cell membrane after passing through the resistor R. U and V represent the resting potential and membrane potential of spiking neurons, respectively. The process of membrane potential accumulation and spiking emission of spiking neurons is shown in Fig. 2. When a neuron receives a spiking signal, its membrane potential will increase. When the membrane potential is greater than the threshold Vth, the neuron emits a spiking signal, and the corresponding membrane potential is reset and is resting.

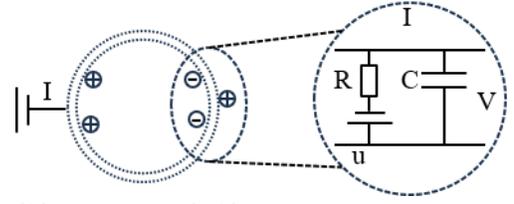

Fig. 1. Spiking neuron switching structure.

Currently, the mainstream neuron models include Integrate-and-fire (IF) model[49], Hodgkin-Huxley (HH) model[50] and LIF model[51]. The LIF neuron model simulates the ion diffusion effect of the battery. When the neuron has no current input, its membrane potential will decay with a time constant. It is known that LIF is the most widely used model to describe neuron dynamics in SNN, which can be expressed by the following formula.

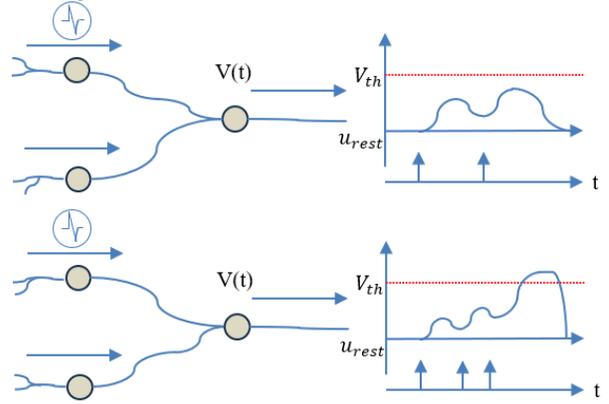

Fig. 2. The spiking accumulation and release process of spiking neurons.

$$\tau \frac{dV(t)}{dt} = -(V(t) - u_{rest}) + RI(t) \quad (1)$$

where $\tau$ is the time constant, $V(t)$ is the neuron membrane potential at time $t$, $u_{rest}$ is the resting potential, $R$ is the input resistance (membrane impedance), and $I(t)$ is the input current. When the membrane potential accumulated by the input current is greater than the threshold, the neuron will release a spiking, and the membrane potential is reset to $u_{rest}$. When the membrane potential is less than the threshold, the neuron does not emit a spiking, and its membrane potential decays to $u_{rest}$ with the time constant $\tau$. As time $t$ changed, the spiking weighted value of the neuron is shown in formula (2):

$$S_j^l(t) = \sum_{i=1}^{n^{l-1}} \omega_{ij}^{l-1} \times x_i^{l-1} \quad (2)$$

Where $S_j^l(t)$ represents the total current accumulated membrane potential of the $j$−th neuron in layer $l$ at time $t$, $n^{l-1}$ represents the number of neurons in layer $l-1$, $\omega_{i(j-1)}^{l-1}$ represents the weight of the synaptic connection between the $i$−th neuron in the $l$ layer and the $j$-th neuron in the $l-1$ layer, $x_i^{l-1}$ represents the total number of spiking events in layer $l-1$ over time $t$, which can be expressed as:

$$x_i^{l-1} = \sum_{k=1}^{t} o_j^{l,t-t_k} \quad (3)$$

Among them, $o_j^{t-t_k}$ represents the moment when the $l$-th layer neuron generates a spiking at $t_k$. The generation of the spiking can be expressed as:

$$o_j^{l,t} = \begin{cases} 0, & v_j^t < V_{th} \\ 1, & v_j^t \geq V_{th} \end{cases} \quad (4)$$

It can be seen that the spiking signal is accumulated in the last layer of the network. When the last time step is reached, the network divides the accumulated membrane potential $V_{mem}$ by the total time step $T_{step}$ to obtain the final output result, which can be expressed as:552

$$Output = \frac{V_{mem}}{T_{step}} \quad (5)$$

*B AAD for backpropagation*

The loss function in this paper is $L$, which is used to reduce the error between the predicted value and the true value. When updating the weight $\omega$ through backpropagation, the following formula can be expressed according to the chain derivation rule:

$$\frac{\partial L}{\partial \omega} = \frac{\partial L}{\partial O} \frac{\partial O}{\partial V} \frac{\partial V}{\partial \omega} \quad (6)$$

Where $L$ is the loss function, $\omega$ is the weight parameter of the network, $O$ is the output spiking after the neuron is activated, and $V$ is the neuron membrane potential. According to the basic principle of spiking emission, we can transform formula (4) into the following unit-step function representation:

$$\Theta(v - v_{th}) = \begin{cases} 1, & (v - v_{th}) \geq 0 \\ 0, & (v - v_{th}) < 0 \end{cases} \quad (7)$$

In the above formula, $v$ is the membrane potential, and $v_{th}$ is the threshold of the release spiking. The derivative of the unit step function is the Dirac function, and its mathematical expression is as follows:

$$\delta(v - v_{th}) = \begin{cases} +\infty, & v = v_{th} \\ 0, & v \neq v_{th} \end{cases} \quad (8)$$

According to the above formula, it can be seen that the Dirac function $\delta(v - v_{th})$ is not differentiable, that is, $\frac{\partial O}{\partial V}$ (the emission of spiking) is not differentiable, resulting in the network being unable to optimize network parameters through the backpropagation algorithm. Therefore, this paper proposes two equivalent differentiable functions (AAD) for the backpropagation of SNN, which optimize network parameters by approximating the Dirac function. The function image is shown in Fig. 3b, and its mathematical representation is shown in formulas (9) and (10):

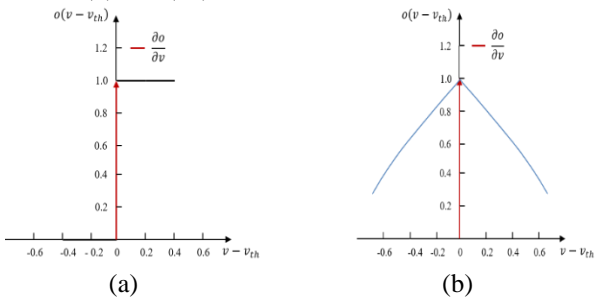

(a)      (b)

Fig. 3. Approximate reciprocal function graph. (a) Primitive functions and Dirac functions. (b) proposed arcsine approximate derivative (AAD).

$$g(v) = |1 - |\arcsin(v)|| < \lambda, \quad \lambda \in (0,1] \quad (9)$$

$$h(v) = \left|1 - \left|\arccos(v) - \frac{\pi}{2}\right|\right| < \lambda, \quad (10)$$
$$\lambda \in (0,1]$$

Among them, $\lambda$ is a custom parameter, $g(v)$ and $h(v)$ are equivalent. If the membrane potential is $v$, and the threshold of the release spiking is $v_{th}$, then the AAD approximate function can be expressed as:

$$\lim_{(v-v_{th}) \to 0^+} g(v - v_{th})$$
$$= \lim_{(v-v_{th}) \to 0^+} \Big|1 \quad (11)$$
$$- |\arcsin(v - v_{th})|\Big| = 1$$

$$\lim_{(v-v_{th}) \to 0^+} h(v - v_{th})$$
$$= \lim_{(v-v_{th}) \to 0^+} \Big|1 \quad (12)$$
$$- \left|\arccos(v - v_{th}) - \frac{\pi}{2}\right|\Big| = 1$$

When the membrane potential is greater than the threshold, the neuron will fire a spiking, which can be expressed as:

$$\lim_{(v-v_{th}) \to 0^+} O(v - v_{th}) = \lim_{(v-v_{th}) \to 0^+} \Theta(v - v_{th}) = 1 \quad (13)$$

$$\lim_{(v-v_{th}) \to 0^+} O(v - v_{th}) = \lim_{(v-v_{th}) \to 0^+} \Theta(v - v_{th})$$
$$= \lim_{(v-v_{th}) \to 0^+} g(v - v_{th}) \quad (14)$$

Obviously, $\frac{\partial g}{\partial V}$, $\frac{\partial h}{\partial V}$, and $\frac{\partial O}{\partial V}$ satisfy the following equation:

$$\frac{\partial O}{\partial V} = \delta(v) \approx \frac{\partial g}{\partial V} \approx \frac{\partial h}{\partial V} \quad (15)$$

Ultimately, network weights $\omega$ The optimization formula (5) can be approximated as:

$$\frac{\partial L}{\partial \omega} = \frac{\partial L}{\partial O} \frac{\partial O}{\partial V} \frac{\partial V}{\partial \omega} = \frac{\partial L}{\partial O} \delta(V) \frac{\partial V}{\partial \omega} \approx \frac{\partial L}{\partial O} \frac{\partial g}{\partial V} \frac{\partial V}{\partial \omega}$$
$$\approx \frac{\partial L}{\partial O} \frac{\partial h}{\partial V} \frac{\partial V}{\partial \omega} \quad (16)$$

*C Spiking mixed convolution (SMC)*

Mixed convolution has attracted much attention from researchers since it was proposed. It can significantly reduce the amounts of parameters required by the network without losing almost any accuracy. Therefore, this paper considers combining mixed convolution with spiking neurons to construct a spiking mixed convolution (SMC). Figure 3 shows the structure diagram of SMC. SMC contains two parts: spiking depth convolution (SDC) and spiking point convolution (SPC). In SMC operation, the input channels are first divided into groups of the same size, and then the SDC performs convolution operations on individual channels of each group using convolution kernels of the same size, storing the values in the back neurons. Subsequently, SPC performs convolution operations on each channel output by the front neurons, and finally, the output of SPC is stored in the back neurons. The generation of spiking sequences in SMC is consistent with Fig. 4.

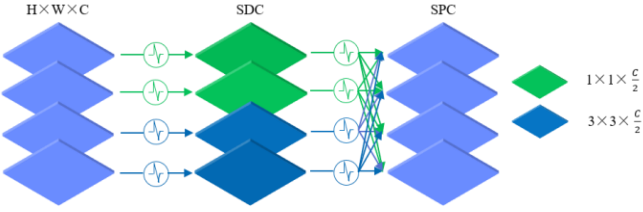

Fig. 4. SMC structure.

A single SDC contains multiple different convolution kernels, and each convolution kernel is connected to a LIF neuron. Assuming that the input is H×W×C, the channels are evenly divided into $g_i$ groups, in a single convolution operation, each group's channel uses different sizes of convolutional kernels $k_i$, in a single convolution operation, then the operation of a single channel in SDC can be expressed as:

$$SDC_{g_{ij}} = LIF\left(k_i(g_{ij})\right) \quad (17)$$

In the above formula, $g_{ij}$ represents the $j$-th channel of the $i$-th groups and $k_i$ represents the convolution kernel of the $i$-th groups. The final result of SDC can be obtained based on the operation of a single channel:

$$SDC = Concat\left(LIF(k_1(g_{11}));\cdots;LIF\left(k_i(g_{ij})\right)\right) \quad (18)$$

Concat represents the fusion operation, and the relationship between $i$ and $j$, and $C$ is $i \times j = C$. After the SDC operation, SPC performs a point convolution operation on each channel. The convolution kernel size is $1 \times 1$, and the input channel and output channel are of the same size. The result after the SPC operation is stored in the posterior neuron. 602

*D Spiking Width Mixed Residual (SWMR) Module*

Generally, deepening or widening a network can significantly improve the feature extraction capability of the network. Currently, the layers of the network can be stacked through residual connections. However, when the network is too deep, the problem of reduced feature reuse will occur and the training time of the network will be significantly increased. Therefore, we consider reducing the depth of the network and widening the network to improve performance. Taking into account the balance between training speed and classification accuracy, the width ratio of the network is very important. Therefore, this paper builds the SWMR module based on SMC, and its network structure is shown in Fig. 5. Compared with the traditional residual module, SWMR adds a new feature extraction branch. Although this will bring some calculations, it can obtain more residual information gain to improve classification accuracy.

Typically, shallower networks contain more detailed information. Therefore, the kernel sizes of the SMC in the two feature extraction branches of the SWMR module in the shallow network are the same, which are $1 \times 1$ and $3 \times 3$. Different from shallow networks, deep networks require richer semantic information. Therefore, the SWMR module adds $5 \times 5$ large kernel convolutions to the deep network. The kernel sizes of the SMC kernels in the two feature extraction branches are $1 \times 1$, $3 \times 3$ and $3 \times 3$, $5 \times 5$ respectively.

The SWMR module can be expressed using the following formula:

$$\begin{aligned}SWMR(X) = X &+ SPC\big(SDC(X)\big) \\ &+ SPC\big(SDC(X)\big)\end{aligned} \quad (19)$$

Among them, $X$ is the input feature map. After combining the above formula with Equation 18, the SWMR module can be expressed as:

$$\begin{aligned}SWMR(X) \\ = X& \\ + SPC&\left(Concat\left(LIF(k_{11}(g_1));LIF(k_{12}(g_2))\right)\right) \\ + SPC&\left(Concat\left(LIF(k_{21}(g_1));LIF(k_{22}(g_2))\right)\right)\end{aligned} \quad (20)$$

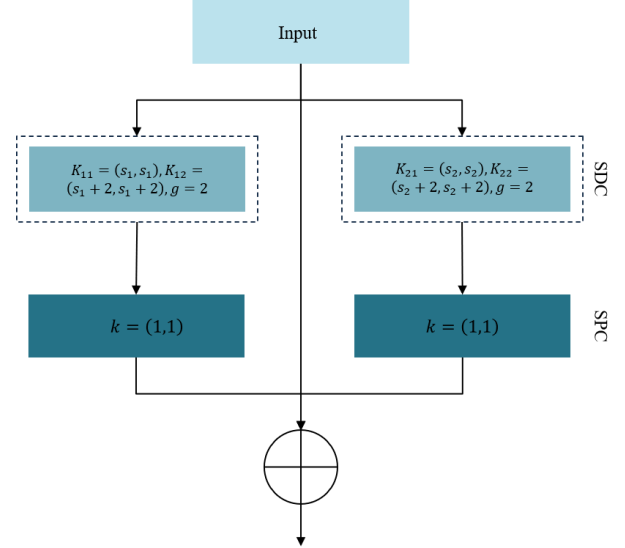

Fig. 5. SWMR module structure.

*E Proposed network*

In SNN, the time step size can greatly affect the performance of the network. A smaller time step size may lead to issues such as loss of detail information and inaccurate network response, affecting the classification accuracy of the network. A larger time step can provide more time for information transmission and processing, which helps the network capture the details and dynamic changes of spiking sequences. However, the time step size is usually positively correlated with the training time, testing time, and complexity of the network. How to achieve higher accuracy under shorter steps is a challenge for HSI classification tasks based on SNN. Therefore, this paper constructs a short-step, high-precision SNN for HSI classification tasks. The overall network structure is shown in Fig. 6, and the specific parameters are shown in TABLE I.

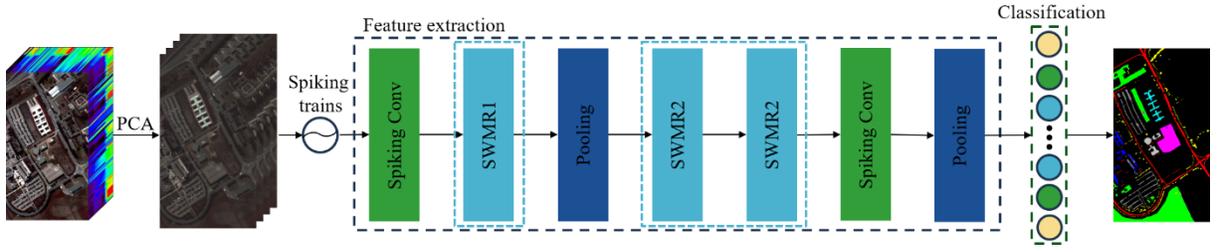

Fig. 6. SNN-SWMR overall network structure diagram.

TABLE I
OVERALL NETWORK PARAMETERS.

| Module | | Operations | Input | Output |
|---|---|---|---|---|
| Feature Extraction | SConv | $k = 3 \times 3, s = 1$ | $17 \times 17, 30$ | $17 \times 17, 64$ |
| | SWMR1 | $k_{11} = 1 \times 1, k_{12} = 3 \times 3, k_{21} = 1 \times 1, k_{22} = 3 \times 3, s = 1$ | $17 \times 17, 64$ | $17 \times 17, 64$ |
| | | $k_1 = 1 \times 1, k_2 = 1 \times 1, s = 1$ | $17 \times 17, 64$ | $17 \times 17, 64$ |
| | Pooling | $k = 2$ | $17 \times 17, 128$ | $8 \times 8, 128$ |
| | SWMR2 | $k_{11} = 1 \times 1, k_{12} = 3 \times 3, k_{21} = 3 \times 3, k_{22} = 5 \times 5, s = 1$ | $8 \times 8, 128$ | $8 \times 8, 128$ |
| | | $k_1 = 1 \times 1, k_2 = 1 \times 1, s = 1$ | $8 \times 8, 128$ | $8 \times 8, 128$ |
| | SWMR2 | $k_{11} = 1 \times 1, k_{12} = 3 \times 3, k_{21} = 3 \times 3, k_{22} = 5 \times 5, s = 1$ | $8 \times 8, 128$ | $8 \times 8, 128$ |
| | | $k_1 = 1 \times 1, k_2 = 1 \times 1, s = 1$ | $8 \times 8, 128$ | $8 \times 8, 128$ |
| | Pooling | $k = 2$ | $8 \times 8, 256$ | $4 \times 4, 256$ |
| Classification | FC | Fully Connect | $4 \times 4, 256$ | Result map |

The network mainly consists of an input layer, a feature extraction layer, and a classification layer. The input layer processes HSI through PCA, extracts effective feature information, and encodes it as spike trains for input into the feature extraction layer. The feature extraction layer includes spiking convolutional layer, pooling layer, and SWMR layer. After receiving the input, the feature extraction layer performs the first feature extraction operation through a spiking convolution with a kernel size of $3 \times 3$, and the output of this spiking convolution will be sent to SWMR1. The output of this convolution will be sent to SWMR1. In SWMR1, $k_1$ and $k_2$ are $1 \times 1$ and $3 \times 3$, respectively. The $1 \times 1$ convolution can transfer more detailed information to the deep network, and the $3 \times 3$ SDC is used for feature extraction. Pooling operation is performed after SWMR1, and the HSI image size is reduced by half. The $k_1$ and $k_2$ of the two residual branches in the deep SWMR2 are $1 \times 1$, $3 \times 3$ and $3 \times 3$, $5 \times 5$ respectively. Using larger convolution kernels in the deep layer can extract more semantic information. Finally, a $1 \times 1$ spiking convolution is used to adjust the dimension, and a pooling operation is performed to adjust the image size. The classification layer is used for the final classification operation and outputs the HSI classification results.

Since the SNN cannot be directly used for training, the training of the SNN in this paper uses the approximate reciprocal algorithm mentioned in Section B for backpropagation to complete the training of the network. The experiments were conducted on six public HSI data sets, and the specific details and results will be elaborated in the experimental section.

EXPERIMENT

*A Dataset description*

This paper uses PU, WHHC, IP, SA, WHLK, and HU data sets to conduct experiments to evaluate the performance of the proposed method. Some data samples in the IP data set are too few, so 80% are used for training and the rest are used as test sets. Except for the IP dataset, 200 samples are randomly selected from the other five datasets for training, and the remaining samples are used for testing.

The PU (Pavia University) dataset was captured by the ROSIS-03 sensor in Germany. The image consists of a total of 115 bands, with a spatial resolution of 1.3m. Among them, 12 bands were removed due to noise, and only 103 bands were retained for subsequent experiments. The dataset download address is:https://www.ehu.eus/ccwintco/index.php/Hyperspectral_Remote_Sensing_Scenes#Pavia_University_scene

The IP (Indian Pines) dataset was captured by AVRIS sensors, and the image contains 220 bands with a spatial resolution of 30m. Among them, 20 bands cannot be reflected by water, so only 200 bands are retained for classification experiments. IP datasets have low spatial resolution and are prone to generating mixed pixels, making classification difficult. The data download address

is:https://www.ehu.eus/ccwintco/index.php/Hyperspectral_Remote_Sensing_Scenes

The WHHC (WHU Hi Han Chuan) dataset and WHLK (WHU Hi Long Kou) dataset were both captured using 8mm focal length head wall nanospectral imaging sensors equipped on the DJI Matrix 600 Pro (DJI M600 Pro) drone platform. WHHC consists of 274 bands and 16 categories. WHLK includes 270 bands and 9 categories. The dataset download address is:http://rsidea.whu.edu.cn/resource_WHUHi_sharing.htm

The HU (Houston 2013) dataset was captured by the ITRES CASI-1500 sensor and consists of 144 bands and 15 categories. The dataset download address is:https://www.grss-ieee.org/community/technical-committees/data-fusion/2013-ieee-grss-data-fusion-contest/

The SA (Salinas Valley) dataset was captured by the AVRIS sensor, and the image contained 224 bands, of which 20 were affected by water vapor absorption, so only 204 bands were retained for classification experiments. The dataset download address is: https://www.ehu.eus/ccwintco/index.php/Hyperspectral_Remote_Sensing_Scenes

*B Experiment setup*

1) Evaluation Metric: Similar to most HSI classification methods, this paper uses three indicators: OA, AA, and Kappa to evaluate the performance of the proposed method. OA refers to the ratio of the total number of correctly classified samples to the total number of all test samples, AA is the average classification accuracy of all categories, and Kappa is used to evaluate the classification consistency of all categories.

2) Implementation Details: In this paper, training samples were randomly selected for each experiment and trained ten times. The initial learning rate is 0.085, and 25 epochs are trained to update the learning rate. The hardware environment for the experiment is: the graphics card is RTX4060Ti, and the CPU is Intel Core i5-12400F. The software environment is Python version 3.8.17, CUDA version 12.2.

*C parameter influence*

1）*The impact of spatial size on classification accuracy.* As is well known, spatial size is crucial for classification accuracy, and either too large or too small spatial size can reduce classification accuracy. Therefore, we consider using time steps as invariants and conducting experiments with different spatial sizes to evaluate their impact on accuracy, selecting the optimal spatial size. The experimental results are shown in TABLE II.

It can be seen that the network proposed in this article significantly improves OA, AA, and Kappa on the four HSI datasets PU, WHHC, WHLK, and HU as the spatial size increases, reaching its maximum at a spatial size of 17. In addition, OA, AA, and Kappa reached their optimal values on the IP and SA HSI datasets at a spatial size of 15, and decreased as the spatial size increased again. Therefore, based on the experimental results, we have selected the optimal spin sizes for PU, IP, WHHC, WHLK, HU, and SA, which are 17, 15, 17, 17, 17, and 17, respectively.

TABLE II
EXPERIMENTAL RESULTS OF DIFFERENT SPATIAL SIZES.

| Dataset | Spatial Size | SNN-SWMR | | |
|---|---|---|---|---|
| | | OA | AA | K×100 |
| PU | 9 | 98.84±0.46 | 98.97±0.46 | 98.45±0.61 |
| | 11 | 99.15±0.38 | 99.21±0.22 | 98.16±0.51 |
| | 13 | 99.38±0.31 | 99.26±0.28 | 99.16±0.41 |
| | 15 | 99.41±0.21 | 99.36±0.21 | 99.20±0.28 |
| | **17** | **99.51±0.23** | **99.41±0.07** | **99.34±0.25** |
| IP | 9 | 98.07±0.44 | 98.95±0.98 | 97.72±0.52 |
| | 11 | 98.44±0.55 | 99.04±0.89 | 98.15±0.65 |
| | 13 | 98.51±0.57 | 99.41±0.30 | 98.24±0.68 |
| | **15** | **98.62±0.29** | **99.38±0.19** | **98.36±0.34** |
| | 17 | 97.77±0.26 | 97.94±020 | 97.39±0.30 |
| WHHC | 9 | 95.58 0.46 | 95.81 0.29 | 94.83 0.54 |
| | 11 | 96.46 0.33 | 96.80 0.26 | 95.85 0.39 |
| | 13 | 97.35±0.33 | 97.59±0.17 | 96.90±0.39 |
| | 15 | 97.50±0.26 | 97.81±0.15 | 97.08±0.30 |
| | **17** | **97.77±0.26** | **97.94±020** | **97.39±0.30** |
| WHLK | 9 | 98.95±0.25 | 98.78±0.37 | 98.62±0.33 |
| | 11 | 98.75±0.59 | 98.81±0.29 | 98.37±0.77 |
| | 13 | 99.04±0.27 | 99.03±0.18 | 98.74±0.35 |
| | 15 | 99.02±0.43 | 98.96±0.24 | 98.72±0.56 |
| | **17** | **99.14±0.33** | **99.01±0.17** | **98.87±0.43** |
| HU | 9 | 99.18±0.20 | 99.35±0.16 | 99.11±0.22 |
| | 11 | 99.35±0.10 | 99.48±0.09 | 99.29±0.11 |
| | 13 | 99.34±0.23 | 99.48±0.19 | 99.29±0.25 |
| | 15 | 99.30±0.24 | 99.43±0.20 | 99.24±0.26 |
| | **17** | **99.46±0.0.21** | **99.55±0.17** | **99.41±0.22** |
| SA | 9 | 99.04±0.41 | 99.62±0.16 | 98.92±0.45 |
| | 11 | 99.14±0.28 | 99.66±0.12 | 99.03±0.31 |
| | 13 | 99.44±0.52 | 99.79±0.16 | 99.38±0.58 |
| | 15 | 99.49±0.23 | 99.79±0.08 | 99.43±0.26 |
| | **17** | **99.55±0.46** | **99.84±0.15** | **99.50±0.51** |

2）*The impact of time steps on classification accuracy.* As shown in the network section proposed by E, the time step size is crucial for classification accuracy. Therefore, after determining the spatial size of the network on each HSI dataset in this paper, the spatial size is taken as the invariant and the time step is taken as the variable to conduct experiments on the impact of time step size on classification accuracy. The experimental results are shown in TABLE III.

TABLE III
EXPERIMENTAL RESULTS AT DIFFERENT TIME STEPS.

| Dataset | Spatial Size | SNN-SWMR | | |
|---|---|---|---|---|
| | | OA | AA | K×100 |
| PU | **10** | **99.51±0.23** | **99.41±0.07** | **99.34±0.25** |
| | 20 | 99.44±0.28 | 99.23±0.38 | 99.25±0.38 |
| | 30 | 99.40±0.72 | 99.26±0.6 | 99.11±0.30 |
| | 40 | 99.31±0.40 | 99.27±0.24 | 99.08±0.53 |
| IP | 10 | 98.62±0.29 | 99.38±0.19 | 98.36±0.34 |
| | **20** | **98.81±0.24** | **99.39±025** | **98.59±0.28** |
| | 30 | 98.40±0.72 | 99.26±0.63 | 98.11±0.85 |
| | 40 | 98.67±0.30 | 99.04±0.87 | 98.43±0.36 |
| WHHC | 10 | 97.77±0.26 | 97.94±020 | 97.39±0.30 |
| | **20** | **97.98±0.29** | **98.21±0.19** | **97.64±0.34** |
| | 30 | 97.78±0.28 | 97.97±0.24 | 97.40±0.33 |
| | 40 | 97.85±0.30 | 98.07±0.25 | 97.48±0.35 |
| WHLK | **10** | **99.14±0.33** | **99.01±0.17** | **98.87±0.43** |
| | 20 | 98.99±0.28 | 98.74±0.28 | 98.67±0.38 |
| | 30 | 98.85±0.44 | 98.55±0.61 | 98.49±0.57 |
| | 40 | 98.65±0.54 | 98.42±0.74 | 98.22±0.74 |
| HU | **10** | **99.43±0.17** | **99.54±0.15** | **99.39±0.19** |
| | 20 | 99.28±0.26 | 99.43±0.20 | 99.21±0.29 |
| | 30 | 99.17±0.16 | 99.33±0.14 | 99.10±0.17 |
| | 40 | 99.15±0.26 | 99.30±0.23 | 99.07±0.28 |
| SA | 10 | 99.55±0.46 | 99.84±0.15 | 99.50±0.51 |
| | **20** | **99.64±0.16** | **99.84±0.06** | **99.59±0.18** |
| | 30 | 99.63±0.23 | 99.84±0.08 | 99.58±0.25 |
| | 40 | 99.53±0.28 | 99.81±0.11 | 99.47±0.31 |

According to TABLE III, it can be seen that the network in this paper only needs 10-time steps to achieve optimal accuracy on the PU, WHLK, and HU HSI datasets, and only needs 20-time steps to achieve optimal accuracy on the IP, WHHC, and SA datasets. The time step experiments on six HSI datasets demonstrate the superiority of the proposed method in terms of short time steps.

RESULT DISCUSSION AND METHOD EVALUATION

*A Comparison with other advanced methods*

Compared with advanced SNN methods such as SNN-SSEM[44] and SNN-DP[45]. Specifically, firstly, the method proposed in this paper is consistent with the SNN-DP and SNN-SSEM methods, except for network models, approximate derivatives, spatial size, time steps, etc. Meanwhile, for the sake of fairness, the spatial size and time step of SNN-DP and SNN-SSEM both adopt the optimal values in their paper. Due to equipment replacement, the accuracy of SNN-DP and SNN-SSEM has slightly fluctuated. In addition, since no experiments were conducted on the WHHC, WHLK, and HU datasets in the SNN-SSEM paper, the spatial size of these three datasets is consistent with this paper.

The classification result images on the PU, IP, WHHC, WHLK, HU, and SA datasets are shown in Fig. 7 to Fig. 12. According to the classification results, SNN-SSEM can provide the clearest classification map and significantly reduce misclassified pixels. The method proposed in this article has some classification errors, while SNN-DP has significant noise.

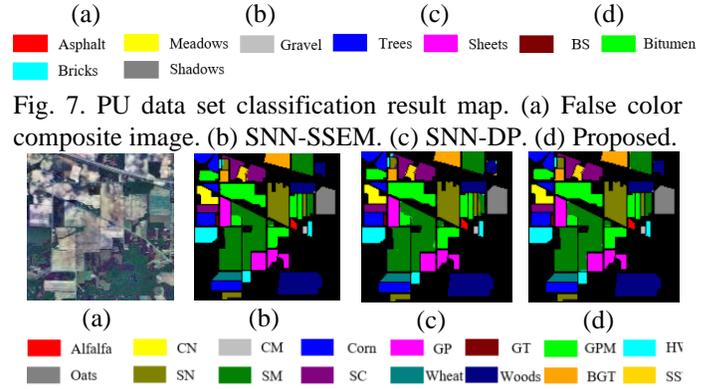

Fig. 7. PU data set classification result map. (a) False color composite image. (b) SNN-SSEM. (c) SNN-DP. (d) Proposed.

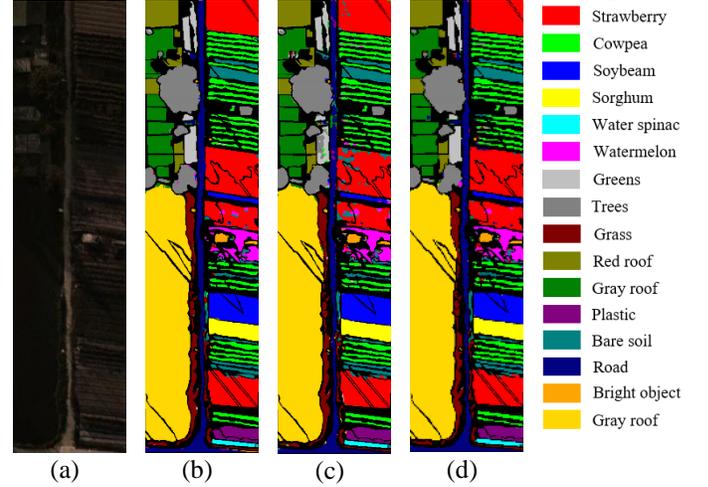

Fig. 8. IP data set classification result map. (a) False color composite image. (b) SNN-SSEM. (c) SNN-DP. (d) Proposed.

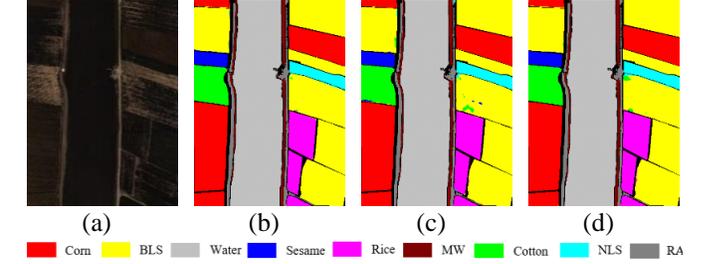

Fig. 9. WHHC data set classification result map. (a) False color composite image. (b) SNN-SSEM. (c) SNN-DP. (d) Proposed.

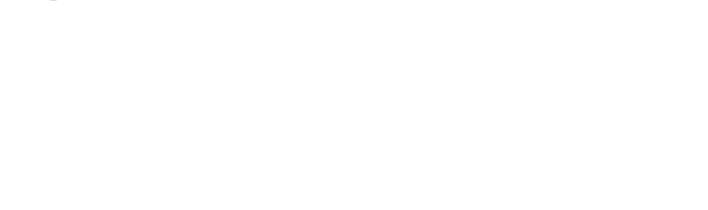

Fig. 10. WHLK data set classification result map. (a) False color composite image. (b) SNN-SSEM. (c) SNN-DP. (d) Proposed.

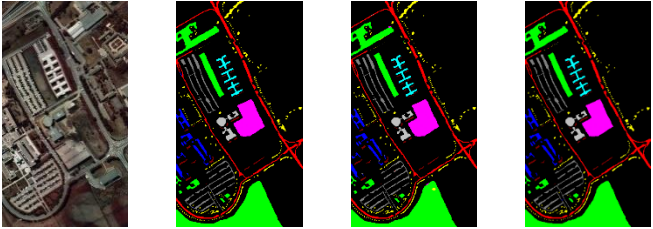

TABLE IV
PU DATA SET COMPARISON RESULTS.

| Class name | SNN-SSEM[44] | SNN-DP[45] | SNN-SWMR (ours) |
|---|---|---|---|
| Asphalt | 99.72 | 99.34 | 99.28 |
| Meadows | 99.47 | 99.30 | 99.74 |
| Gravel | 99.18 | 98.72 | 99.65 |
| Tress | 99.48 | 98.12 | 97.85 |
| Painted metal sheets | 100.00 | 99.86 | 99.84 |
| Bare Soil | 99.85 | 99.98 | 99.99 |
| Bitumen | 100.00 | 99.96 | 100.00 |
| Self-Blocking Bricks | 99.66 | 98.78 | 99.06 |
| Shadows | 99.84 | 99.69 | 99.30 |
| OA | **99.59±0.31** | 99.30±0.11 | **99.51±0.23** |
| AA | **99.69±0.11** | 99.26±0.18 | **99.41±0.07** |
| Kappa | **99.45±0.42** | 99.06±0.14 | **99.34±0.25** |
| Time step | 70 | 40 | **10** |
| Train time | 636.66 | 462.32 | **237.62** |
| Test time | 88.59 | 45.75 | **28.32** |

TABLE V
IP DATA SET COMPARISON RESULTS.

| Class name | SNN-SSEM[44] | SNN-DP[45] | SNN-SWMR (ours) |
|---|---|---|---|
| Alfalfa | 100.00 | 98.89 | 98.89 |
| Corn-notill | 98.02 | 96.48 | 97.84 |
| Corn-mintill | 99.65 | 99.02 | 99.43 |
| Corn | 100.00 | 100.00 | 100.00 |
| Grass-pasture | 99.79 | 99.33 | 99.61 |
| Grass-tress | 99.89 | 99.21 | 99.64 |
| Grass-pasture-mowed | 100.00 | 100.00 | 100.00 |
| Hay-windrowed | 100.00 | 100.00 | 100.00 |
| Oats | 100.00 | 100.00 | 100.00 |
| Soybean-notill | 99.34 | 98.06 | 98.99 |
| Soybean-mintill | 97.77 | 96.99 | 98.07 |
| Soybean-clean | 98.91 | 98.45 | 99.21 |
| Wheat | 100.00 | 100.00 | 100.00 |
| Woods | 99.69 | 99.12 | 99.69 |
| Buildings-Grass-Trees-Drives | 100.00 | 99.89 | 99.95 |
| Stone-Steel-Towers | 99.44 | 99.44 | 98.89 |
| OA | **98.82±0.24** | 98.00±0.44 | 98.81±0.24 |
| AA | **98.53±0.15** | 99.05±0.36 | 99.39±0.25 |
| Kappa | **98.60±0.28** | 97.62±0.52 | 98.59±0.28 |
| Time step | 100 | 30 | **20** |
| Train time | 1194.15 | 624.77 | **530.69** |
| Test time | 21.54 | 8.44 | **7.98** |

TABLE VI
WHHC DATA SET COMPARISON RESULTS.

| Class name | SNN-SSEM[44] | SNN-DP[45] | SNN-SWMR (ours) |
|---|---|---|---|
| Strawberry | 96.25 | 96.13 | 97.20 |
| Cowpea | 97.74 | 94.59 | 96.40 |
| Soybeam | 99.30 | 98.90 | 99.37 |
| Sorghum | 99.96 | 99.83 | 99.74 |
| Water spinach | 100.00 | 100.00 | 100.00 |
| Watermelon | 97.92 | 97.10 | 98.11 |
| Greens | 99.40 | 98.78 | 99.01 |
| Trees | 95.45 | 95.76 | 96.73 |
| Grass | 98.10 | 96.52 | 98.04 |
| Red roof | 99.32 | 98.81 | 99.00 |
| Gray roof | 98.95 | 98.63 | 98.92 |
| Plastic | 99.86 | 99.68 | 99.85 |
| Bare soil | 93.47 | 91.13 | 93.29 |
| Road | 97.01 | 95.89 | 97.28 |
| Bright object | 99.56 | 99.31 | 99.38 |
| Water | 99.1 | 99.29 | 99.07 |
| OA | 97.87±0.28 | 97.33±0.47 | **97.98±0.29** |
| AA | **98.21±0.18** | 97.52±0.50 | 98.21±0.19 |
| Kappa | 97.50±0.33 | 96.88±0.55 | **97.64±0.34** |
| Time step | 100 | 40 | **20** |
| Train time | 1839.67 | 1106.46 | **746.65** |
| Test time | 921.41 | 385.13 | **287.15** |

TABLE VII
WHLK Data Set Comparison Results.

| Class name | SNN-SSEM[44] | SNN-DP[45] | SNN-SWMR (ours) |
|---|---|---|---|
| Corn | 99.87 | 99.86 | 99.69 |
| Cotton | 99.87 | 99.19 | 99.53 |
| Seasame | 99.99 | 99.87 | 99.77 |
| Broad-leaf soybean | 98.17 | 98.40 | 98.84 |
| Narrow-leaf soybean | 99.95 | 99.57 | 99.38 |
| Rice | 99.77 | 99.59 | 99.36 |
| Water | 99.57 | 99.32 | 99.30 |
| Roads and houses | 97.63 | 97.53 | 97.24 |
| Mixed weed | 98.58 | 97.80 | 97.97 |
| OA | 99.13±0.19 | 99.05±0.34 | **99.14±0.33** |
| AA | **99.27±0.10** | 99.01±0.16 | 99.01±0.17 |
| Kappa | 98.86±0.25 | 98.75±0.45 | **98.87±0.43** |
| Time step | 100 | 30 | **10** |
| Train time | 1028.12 | 396.15 | **244.44** |
| Test time | 709.73 | 192.93 | **144.13** |

TABLE VIII
HU Data Set Comparison Results

| Class name | SNN-SSEM[44] | SNN-DP[45] | SNN-SWMR (ours) |
|---|---|---|---|
| Healthy grass | 99.35 | 99.01 | 99.14 |
| Stressed grass | 99.88 | 99.54 | 99.77 |
| Synthetic grass | 99.96 | 99.66 | 99.70 |
| Trees | 99.56 | 99.63 | 99.33 |
| Soil | 100.00 | 99.91 | 99.94 |
| Water | 100.00 | 100.00 | 100.00 |
| Residential | 99.45 | 98.60 | 99.26 |
| Commercial | 97.79 | 98.03 | 98.16 |
| Road | 99.28 | 98.47 | 99.25 |
| Highway | 99.92 | 99.91 | 99.98 |
| BRailway | 99.74 | 99.76 | 99.84 |
| Parking Lot 1 | 99.06 | 99.09 | 99.05 |
| Parking Lot 2 | 99.81 | 99.55 | 99.70 |
| Tennis Court | 100.00 | 100.00 | 100.00 |
| Running Track | 100.00 | 100.00 | 100.00 |
| OA | **99.47±0.13** | 99.20±0.15 | 99.43±0.17 |
| AA | **99.59±0.10** | 99.36±0.12 | 99.54±0.15 |
| Kappa | **99.43±0.14** | 99.13±0.17 | 99.39±0.19 |
| Time step | 100 | 30 | **10** |
| Train time | 1698.99 | 520.79 | **387.52** |
| Test time | 42.80 | 10.57 | **8.28** |

TABLE IX
SA Data Set Comparison Results.

| Class name | SNN-SSEM[44] | SNN-DP[45] | SNN-SWMR (ours) |
|---|---|---|---|
| Brocoli_green_weeds_1 | 100.00 | 99.99 | 99.99 |
| Brocoli_green_weeds_2 | 100.00 | 99.98 | 99.99 |
| Fallow | 100.00 | 100.00 | 99.98 |
| Fallow_arough_plow | 99.96 | 99.93 | 99.97 |
| Fallow_smooth | 99.76 | 99.79 | 99.60 |
| Stubble | 100.00 | 99.99 | 99.99 |
| Celery | 100.00 | 99.98 | 99.96 |
| Grapes_untrained | 99.05 | 98.80 | 98.93 |
| Soli_vinyard_develop | 100.00 | 100.00 | 100.00 |
| Corn_senesced_treen_weeds | 99.97 | 99.70 | 99.86 |
| Lettuce_romaine_4wk | 100.00 | 100.00 | 99.98 |
| Lettuce_romaine_5wk | 100.00 | 100.00 | 99.93 |
| Lettuce_romaine_6mk | 99.99 | 99.99 | 99.99 |
| Lettuce_romaine_7mk | 100.00 | 99.93 | 99.98 |
| Vinyard_untrained | 99.40 | 99.34 | 99.33 |
| Vinyard_vertical_trellis | 100.00 | 99.93 | 99.96 |
| OA | **99.70±0.17** | 99.61±0.25 | 99.64±0.16 |
| AA | **99.88±0.07** | 99.84±0.08 | 99.84±0.06 |
| Kappa | **99.66±0.19** | 99.56±0.28 | 99.59±0.18 |
| Time step | 100 | 40 | **20** |
| Train time | 1506.46 | 1121.39 | **768.55** |
| Test time | 146.82 | 78.23 | **57.74** |

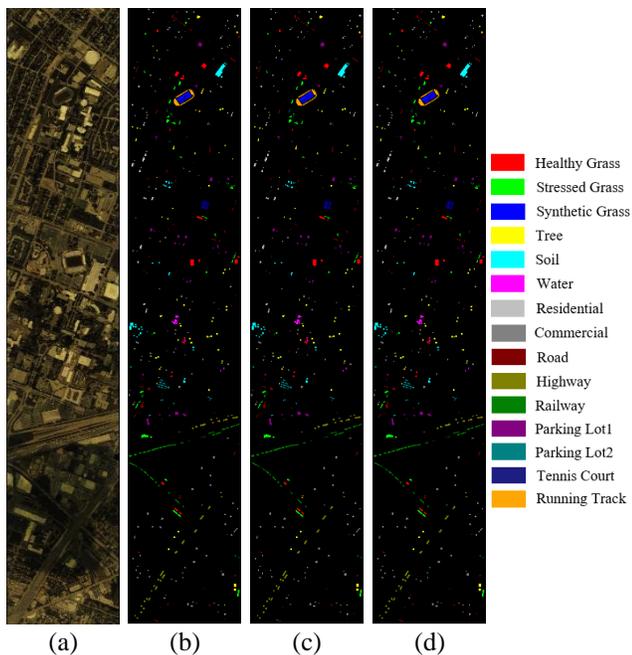

(a) (b) (c) (d)

Fig. 11. HU data set classification result map. (a) False color composite image. (b) SNN-SSEM. (c) SNN-DP. (d) Proposed.

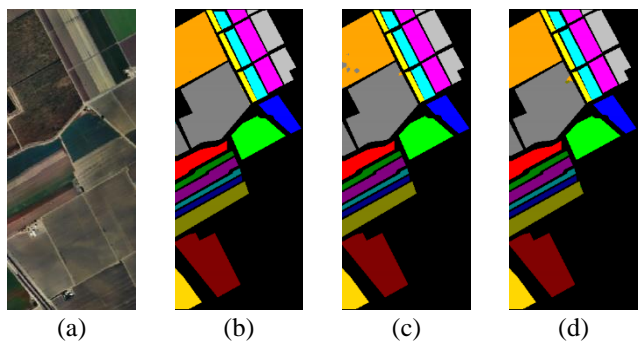

(a) (b) (c) (d)

Fig. 12. SA data set classification result map. (a) False color composite image. (b) SNN-SSEM. (c) SNN-DP. (d) Proposed.

The comparative data of experimental results on PU, IP, WHHC, WHLK, HU, and SA datasets are shown in TABLE IV-TABLE IX. Obviously, from the perspective of classification accuracy, the SNN-SSEM algorithm has a slight advantage over SNN-SWMR in terms of classification accuracy. The two are at the same level, while the SNN-DP classification accuracy is relatively low. From the perspectives of time step size, training time, and testing time, the algorithm proposed in this paper has absolute advantages. Compared with SNN-SSEM, the time step for SNN-SWMR to achieve optimal accuracy has been reduced by approximately 84% (86%, 80%, 80%, 90%, 90%, and 80% for the six datasets, respectively), and the training and testing time has been reduced by an average of about 63% (61%, 56%, 59%, 76%, 77%, and 49% for the six datasets, respectively) and 70% (68%, 63%, 69%, 80%, 81%, and 61% for the six datasets, respectively). Compared with SNN-DP, the time step for SNN-SWMR to achieve optimal accuracy has been reduced by 57% (75%, 33%, 50%, 67%, 67%, and 50% respectively for the six datasets), and the training and testing time has been reduced by an average of about 32% (49%, 15%, 33%, 38%, 26%, and 31%) and 23% (38%, 4%, 25%, 25%, 22%, and 26%).

*B Effect of Approximate Derivatives*

To analyze the impact of the approximate derivatives proposed in this article on network performance improvement, we conducted experiments on six public datasets to evaluate the approximate derivatives proposed in this paper. For the sake of fairness, we conducted experiments on the SNN-DP network using the square approximate derivative (SAD) [45] and the arcsine approximate derivative (AAD) proposed in this paper. The experimental results are shown in TABLE X. SNN-DP adopts the optimal spatial size in the original paper on PU, IP, WHHC, WHLK, HU, and SA datasets, with values of 13, 13, 17, 11, 17, and 15, respectively. Obviously, after replacing the approximate derivative with AAD, except for the SA dataset, the classification performance has advantages at multiple time steps across multiple datasets, achieving a maximum accuracy improvement of about 1% when only replacing the approximate reciprocal.

*C The Influence of Convolutional Kernel Size on SMC*

As we all know, the convolution kernel size is crucial to the network. Most current networks use 3 × 3 convolution kernels to reduce the amount of network parameters and calculations.

However, using appropriate convolutional kernels at appropriate locations may have an advantage in classification accuracy over a network with a single kernel size. Therefore, this paper compares the impact of the convolution kernel size on the network in SMC. The experimental results are shown in TABLE XI. SNN-SWMR (1,3) means that all SMC in the network only use two kernel sizes: 1×1 and 3×3. SNN-SWMR (3,5) means that all SMC in the network use only two kernel sizes: 1×1 and 3×3. It can be seen that as the convolution kernel increases, the training time and testing time of the network increase. However, OA, AA, and Kappa are not positively related to the size of the convolution kernel. SNN-SWMR (1,3) has the shortest training time and testing time, but its classification effect on the six data sets is the worst. SNN-SWMR (3,5) takes the longest training and testing time,

TABLE X
COMPARATIVE RESULTS OF AAD AND SAD APPROXIMATE DERIVATIVES.

| Dataset | Time Steps | SNN-DP+SAD | | | SNN-DP+AAD | | |
|---|---|---|---|---|---|---|---|
| | | OA | AA | K×100 | OA | AA | K×100 |
| PU | 10 | 99.03±0.43 | 99.08±0.27 | 98.70±0.58 | **99.20±0.27** | **99.24±0.22** | **98.93±0.37** |
| | 20 | 99.19±0.21 | 99.20±0.16 | 99.91±0.29 | **99.46±0.11** | **99.45±0.11** | **99.28±0.15** |
| | 30 | 99.28±0.20 | 99.24±0.14 | 99.03±0.27 | **99.48±0.22** | **99.49±0.19** | **99.30±0.30** |
| | 40 | 99.30±0.11 | 99.26±0.18 | 99.06±0.14 | **99.47±0.16** | **99.45±0.12** | **99.29±0.21** |
| IP | 10 | 97.44±0.46 | 99.01±0.33 | 97.50±0.54 | **98.45±0.46** | **99.05±0.92** | **98.16±0.54** |
| | 20 | 97.93±0.44 | 98.90±0.51 | 97.55±0.52 | **98.59±0.39** | **99.31±0.22** | **98.32±0.46** |
| | 30 | 98.00±0.44 | 99.05±0.36 | 97.62±0.52 | **98.58±0.47** | **99.08±1.16** | **98.32±0.55** |
| | 40 | 97.79±0.68 | 98.95±0.45 | 97.38±0.80 | **98.59±0.38** | **99.30±0.53** | **98.33±0.45** |
| WHHC | 10 | 96.88±0.29 | 97.20±0.15 | 96.36±0.34 | **97.47±0.44** | **97.80±0.26** | **97.04±0.52** |
| | 20 | 97.14±0.38 | 97.44±0.30 | 96.65±0.44 | **97.71±0.23** | **97.97±0.18** | **97.32±0.27** |
| | 30 | 97.32±0.33 | 97.57±0.33 | 96.87±0.39 | **97.71±0.26** | **97.94±0.24** | **97.32±0.30** |
| | 40 | 97.33±0.47 | 97.52±0.50 | 96.88±0.55 | **97.44±0.26** | **97.61±0.23** | **97.00±0.30** |
| WHLK | 10 | 98.88±0.35 | 98.83±0.35 | 98.53±0.45 | **99.04±0.30** | **99.06±0.21** | **98.74±0.40** |
| | 20 | 98.94±0.39 | 98.94±0.27 | 98.61±0.51 | **99.01±0.35** | **99.09±0.18** | **98.70±0.46** |
| | 30 | **99.05±0.34** | 99.01±0.16 | 98.75±0.45 | 99.04±0.45 | **99.10±0.24** | **98.73±0.58** |
| | 40 | 98.97±0.45 | 98.98±0.24 | 98.64±0.58 | **99.15±0.27** | **99.13±0.17** | **98.88±0.36** |
| HU | 10 | 99.05±0.25 | 99.24±0.21 | 98.96±0.27 | **99.33±0.22** | **99.46±0.17** | **99.27±0.24** |
| | 20 | 99.11±0.42 | 99.28±0.31 | 99.03±0.46 | **99.35±0.19** | **99.48±0.17** | **99.29±0.21** |
| | 30 | 99.20±0.15 | 99.36±0.12 | 99.13±0.17 | **99.30±0.30** | **99.44±0.25** | **99.24±0.32** |
| | 40 | 99.07±0.19 | 99.26±0.16 | 98.99±0.21 | **99.34±0.17** | **99.48±0.12** | **99.28±0.28** |
| SA | 10 | 99.39±0.37 | 99.78±0.11 | 99.31±0.42 | **99.53±0.29** | **99.81±0.11** | **99.48±0.32** |
| | 20 | 99.46±0.20 | 99.78±0.05 | 99.46±0.20 | **99.46±0.28** | **99.81±0.09** | **99.40±0.31** |
| | 30 | 99.60±0.25 | 99.84±0.09 | 99.60±0.25 | **99.60±0.19** | **99.84±0.09** | **99.55±0.22** |
| | 40 | **99.63±0.17** | **99.84±0.08** | **99.59±0.19** | 99.56±0.27 | 99.84±0.11 | 99.51±0.30 |

TABLE XI
COMPARATIVE RESULTS OF DIFFERENT CONVOLUTION KERNELS IN SMC.

| Dataset | Model | Train Time(s) | Test Time(s) | OA | AA | K×100 |
|---|---|---|---|---|---|---|
| PU | SNN-SWMR (ours) | 237.62 | 28.32 | **99.51±0.23** | **99.41±0.07** | **99.34±0.25** |
| | SNN-SWMR (1,3) | **234.25** | **28.14** | 99.33±0.22 | 99.20±0.23 | 99.10±0.30 |
| | SNN-SWMR (3,5) | 270.27 | 30.19 | 99.45±0.24 | 99.33±0.27 | 99.27±0.33 |
| IP | SNN-SWMR (ours) | 530.69 | 7.98 | **98.81±0.24** | **99.39±025** | **98.59±0.28** |
| | SNN-SWMR (1,3) | **519.44** | **7.96** | 98.44±0.28 | 99.37±0.12 | 98.15±0.33 |
| | SNN-SWMR (3,5) | 626.66 | 8.77 | 98.62±0.34 | 98.96±0.97 | 98.37±0.40 |
| WHHC | SNN-SWMR (ours) | 746.65 | 287.15 | 97.98±0.29 | 98.21±0.19 | 97.64±0.34 |
| | SNN-SWMR (1,3) | **737.81** | **293.36** | 97.76±0.36 | 97.96±0.29 | 97.38±0.42 |
| | SNN-SWMR (3,5) | 868.40 | 315.84 | **98.10±0.25** | **98.24±0.23** | **98.78±0.29** |
| WHLK | SNN-SWMR (ours) | 244.44 | 144.13 | **99.14±0.33** | 99.01±0.17 | **98.87±0.43** |

| Dataset | Model | Train Time(s) | Test Time(s) | OA | AA | K×100 |
|---|---|---|---|---|---|---|
| | SNN-SWMR (1,3) | **235.26** | **141.22** | 99.04±0.18 | 98.92±0.50 | 98.74±0.63 |
| | SNN-SWMR (3,5) | 273.50 | 152.19 | 99.12±0.26 | 99.00±0.21 | 99.84±0.34 |
| HU | SNN-SWMR (ours) | 387.52 | 8.28 | **99.43±0.17** | **99.54±0.15** | **99.39±0.19** |
| | SNN-SWMR (1,3) | **357.39** | **7.80** | 99.30±0.18 | 99.44±0.14 | 99.24±0.19 |
| | SNN-SWMR (3,5) | 434.83 | 8.92 | 99.33±0.19 | 99.47±0.14 | 99.27±0.21 |
| | SNN-SWMR (ours) | 768.55 | 57.74 | **99.64±0.16** | **99.84±0.06** | **99.59±0.18** |
| SA | SNN-SWMR (1,3) | **637.69** | **52.68** | 99.48±**0.37** | 99.79±**0.13** | 99.41±**0.42** |
| | SNN-SWMR (3,5) | 866.20 | 57.48 | 99.56±0.31 | 99.82±0.12 | 99.51±0.35 |

TABLE XII
COMPARISON RESULTS OF SWMR MODULES WITH DIFFERENT WIDTH FACTORS.

| Dataset | Model | Train Time(s) | Test Time(s) | OA | AA | K×100 |
|---|---|---|---|---|---|---|
| | SNN-SWMR (ours) | 237.62 | 28.32 | **99.51±0.23** | **99.41±0.07** | **99.34±0.25** |
| PU | SNN-SMR | **158.80** | **19.62** | 99.33±0.37 | 99.34±0.16 | 99.10±0.49 |
| | SNN-3SWMR | 339.28 | 38.37 | 99.47±0.34 | 99.37±0.25 | 99.28±0.46 |
| | SNN-SWMR (ours) | 530.69 | 7.98 | **98.81±0.24** | 99.39±025 | **98.59±0.28** |
| IP | SNN-SMR | **346.27** | **5.27** | 98.52±0.32 | 99.34±0.17 | 98.25±0.37 |
| | SNN-3SWMR | 725.74 | 10.78 | 98.52±0.41 | **98.43±0.16** | 98.24±0.48 |
| | SNN-SWMR (ours) | 746.65 | 287.15 | **97.98±0.29** | **98.21±0.19** | **97.64±0.34** |
| WHHC | SNN-SMR | **489.50** | **195.78** | 97.92±0.33 | 98.16±0.24 | 97.57±0.39 |
| | SNN-3SWMR | 1024.17 | 383.12 | 97.95±0.29 | 98.16±0.23 | 97.59±0.34 |
| | SNN-SWMR (ours) | 244.44 | 144.13 | **99.14±0.33** | **99.01±0.17** | **98.87±0.43** |
| WHLK | SNN-SMR | **158.66** | **96.47** | 99.00±0.37 | 99.01±0.22 | 98.69±0.49 |
| | SNN-3SWMR | 334.06 | 187.00 | 99.10±0.26 | 98.98±0.15 | 98.82±0.34 |
| | SNN- SWMR (ours) | 387.52 | 8.28 | 99.43±0.17 | 99.54±0.15 | 99.39±0.19 |
| HU | SNN-SMR | **259.04** | **5.87** | 99.33±0.14 | 99.47±0.11 | 99.27±0.15 |
| | SNN-3 SWMR | 537.24 | 11.22 | **99.46±0.21** | **99.55±0.17** | **99.41±0.22** |
| | SNN-SWMR (ours) | 768.55 | 57.74 | **99.64±0.16** | 99.84±0.06 | **99.59±0.18** |
| SA | SNN-SMR | **422.86** | **34.86** | 99.56±0.26 | 99.83±0.08 | 99.50±0.29 |
| | SNN-3SWMR | 911.59 | 72.49 | 99.62±0.23 | **99.85±0.07** | 99.58±0.25 |

but its classification effect is only slightly better than SNN-SWMR on the WHHC data set. SNN-SWMR outperformed SNN SWMR (1,3) in classification performance on all six datasets, with small increases in training and testing time. At the same time, SNN SWMR outperformed SNN SWMR (3,5) in classification performance on PU, IP, WHHC, WHLK, SA, and HU datasets, and significantly reduced training and testing time (12%, 15%, 14%, 11%, 11%, 11%). According to the experimental results and analysis, the convolutional kernel configuration of SNN-SWMR can achieve a certain balance in classification accuracy, training time, and testing time.

*D Effect of SWMR module*

To compare the impact of the width factor on classification performance in the SWMR module, this paper uses three spiking residual modules with a width factor of 1 (SMR), a width factor of 2 (SWMR), and a width factor of 3 (3SWMR). The experimental results are shown in TABLE XII. According to the experimental results, it can be seen that a limited increase in the width factor (that is, network width) can improve the classification performance of the network, but an increase in network width can lead to an increase in training and testing time. At the same time, when the width factor is 3 (that is, SNN-3SWMR), the improvement of the network classification performance can be almost negligible (such as the OA, AA, and Kappa of WHHC and HU datasets only improving by 0.03%, 0.05%, 0.05% and 0.03%, 0.01%, 0.02%), and even the classification performance deteriorates (such as PU, IP, SA, WHLK). Therefore, increasing the width of the network appropriately can improve its performance. Considering the balance between training time, testing time, and classification performance, the width factor of the SNN-SWMR network in this paper is 2.

CONCLUSION

This paper constructs a SNN based on LIF neurons that can be directly trained. The SNN network consists of spiking convolution layer, SMWR module, pooling layer, and classification layer, a7nd is directly trained through back propagation using the AAD designed in this paper. This paper conducted experimental evaluations on six publicly available HSI datasets, and SNN-SWRB achieved good results, achieving optimal accuracy in as little as 10-time steps. Specifically, compared to the advanced spiking neural networks SNN-SSEM and SNN-DP in the same category, the optimal model has similar classification performance. The algorithm proposed in this paper has reduced the time step, training time, and testing time by about 84%, 63%, and 70% compared to SNN-SSEM under the optimal model, and by about 57%, 32%, and 23% compared to SNN-DP. Experiments have shown that the proposed method can achieve high accuracy in short time steps, solving the problem of long training and testing time for SNN algorithms. It is of great significance for promoting the application of SNN based

HSI classification algorithms in unmanned autonomous devices. It is of great significance to promote the practical application of SNN based HSI classification algorithms in unmanned autonomous devices such as spaceborne and airborne devices. However, the effectiveness of SNN-SWMR is limited when there are very few samples, and there are usually fewer available samples in HSI. Therefore, next we will consider how to train and infer HSI classification algorithms in unmanned autonomous devices such as spaceborne and airborne systems with limited available samples.

REFERENCES

[1] S. Zhang, J. Li, Z. Wu, and A. Plaza, "Spatial discontinuity-weighted sparse unmixing of hyperspectral images," *IEEE Trans. Geosci. Remote Sens.,* vol. 56, no. 10, pp. 5767-5779, 2018.

[2] D. Hong, J. Yao, D. Meng, Z. Xu, and J. Chanussot, "Multimodal GANs: Toward crossmodal hyperspectral–multispectral image segmentation," *IEEE Trans. Geosci. Remote Sens.,* vol. 59, no. 6, pp. 5103-5113, 2020.

[3] X. Yang and Y. Yu, "Estimating soil salinity under various moisture conditions: An experimental study," *IEEE Trans. Geosci. Remote Sens.,* vol. 55, no. 5, pp. 2525-2533, 2017.

[4] B. Zhang, L. Zhao, and X. Zhang, "Three-dimensional convolutional neural network model for tree species classification using airborne hyperspectral images," *Remote Sens Environ.,* vol. 247, p. 111938, 2020.

[5] L. Liang *et al.*, "Estimation of crop LAI using hyperspectral vegetation indices and a hybrid inversion method," *Remote Sens Environ.,* vol. 165, pp. 123-134, 2015.

[6] Q. Li, F. K. K. Wong, and T. Fung, "Mapping multi-layered mangroves from multispectral, hyperspectral, and LiDAR data," *Remote Sens Environ.,* vol. 258, p. 112403, 2021.

[7] F. Melgani and L. Bruzzone, "Classification of hyperspectral remote sensing images with support vector machines," *IEEE Trans. Geosci. Remote Sens.,* vol. 42, no. 8, pp. 1778-1790, 2004.

[8] J. Li, J. M. Bioucas-Dias, and A. Plaza, "Semisupervised hyperspectral image segmentation using multinomial logistic regression with active learning," *IEEE Trans. Geosci. Remote Sens.,* vol. 48, no. 11, pp. 4085-4098, 2010.

[9] B. Du and L. Zhang, "Target detection based on a dynamic subspace," *Pattern Recognit.,* vol. 47, no. 1, pp. 344-358, 2014.

[10] J. Li, J. M. Bioucas-Dias, and A. Plaza, "Spectral–spatial hyperspectral image segmentation using subspace multinomial logistic regression and Markov random fields," *IEEE Trans. Geosci. Remote Sens.,* vol. 50, no. 3, pp. 809-823, 2011.

[11] A. Fu, X. Ma, and H. Wang, "Classification of hyperspectral image based on hybrid neural networks," in *Proc. IEEE Int. Geosci. Remote Sens. Symp*, 2018: IEEE, pp. 2643-2646.

[12] G. Hughes, "On the mean accuracy of statistical pattern recognizers," *IEEE Trans. Inf Theory,* vol. 14, no. 1, pp. 55-63, 1968.

[13] Q. Du, "Modified Fisher's linear discriminant analysis for hyperspectral imagery," *IEEE Geosci. Remote Sens. Lett.,* vol. 4, no. 4, pp. 503-507, 2007.

[14] P. Deepa and K. Thilagavathi, "Feature extraction of hyperspectral image using principal component analysis and folded-principal component analysis," in *Int. Conf. Electron. Commun. Syst., ICECS*, 2015: IEEE, pp. 656-660.

[15] M. Imani and H. Ghassemian, "Principal component discriminant analysis for feature extraction and classification of hyperspectral images," in *Iran. Conf. Intelligent Syst., ICIS*, 2014: IEEE, pp. 1-5.

[16] C. Jayaprakash, B. B. Damodaran, V. Sowmya, and K. Soman, "Dimensionality reduction of hyperspectral images for classification using randomized independent component analysis," in *Int. Conf. Signal Process. Integr. Networks, SPIN*, 2018: IEEE, pp. 492-496.

[17] J. Wang and C.-I. Chang, "Independent component analysis-based dimensionality reduction with applications in hyperspectral image analysis," *IEEE Trans. Geosci. Remote Sens.,* vol. 44, no. 6, pp. 1586-1600, 2006.

[18] X. Li *et al.*, "Multi-view learning for hyperspectral image classification: An overview," *Neurocomputing,* vol. 500, pp. 499-517, 2022.

[19] J. Wang, G. Zhang, M. Cao, and N. Jiang, "Semi-supervised classification of hyperspectral image based on spectral and extended morphological profiles," in *Workshop Hyperspectral Image Signal Proces.: Evol. Remote Sens.*, 2016: IEEE, pp. 1-4.

[20] X. Zhang and W. Qi, "Hyperspectral image classification based on extended morphological attribute profiles and abundance information," in *2018 9th Workshop on Hyperspectral Image and Signal Processing: Evolution in Remote Sensing (WHISPERS)*, 2018: IEEE, pp. 1-5.

[21] L. Fang, S. Li, X. Kang, and J. A. Benediktsson, "Spectral–spatial hyperspectral image classification via multiscale adaptive sparse representation," *IEEE Trans. Geosci. Remote Sens.,* vol. 52, no. 12, pp. 7738-7749, 2014.

[22] Y. Chen, Z. Lin, X. Zhao, G. Wang, and Y. Gu, "Deep learning-based classification of hyperspectral data," *IEEE J. Sel.Topics Appl. Earth Observ. Remote Sens.,* vol. 7, no. 6, pp. 2094-2107, 2014.

[23] T. Li, J. Zhang, and Y. Zhang, "Classification of hyperspectral image based on deep belief networks," in *IEEE Int. Conf. Image Process., ICIP*, 2014: IEEE, pp. 5132-5136.

[24] Y. Chen, H. Jiang, C. Li, X. Jia, and P. Ghamisi, "Deep feature extraction and classification of hyperspectral images based on convolutional neural networks," *IEEE Trans. Geosci. Remote Sens.,* vol. 54, no. 10, pp. 6232-6251, 2016.

[25] W. Li, G. Wu, F. Zhang, and Q. Du, "Hyperspectral image classification using deep pixel-pair features," *IEEE Trans. Geosci. Remote Sens.,* vol. 55, no. 2, pp. 844-853, 2016.

[26] X. Liu, C. Leng, X. Niu, Z. Pei, I. Cheng, and A. Basu, "Find Small Objects in UAV Images by Feature Mining

[27] K. He, X. Zhang, S. Ren, and J. Sun, "Deep residual learning for image recognition," in *Proc IEEE Comput Soc Conf Comput Vision Pattern Recognit*, 2016, pp. 770-778.

[28] H. Gao, Y. Yang, C. Li, L. Gao, and B. Zhang, "Multiscale residual network with mixed depthwise convolution for hyperspectral image classification," *IEEE Trans. Geosci. Remote Sens.,* vol. 59, no. 4, pp. 3396-3408, 2020.

[29] Y. Li, Q. Xu, W. Li, and J. Nie, "Automatic clustering-based two-branch CNN for hyperspectral image classification," *IEEE Trans. Geosci. Remote Sens.,* vol. 59, no. 9, pp. 7803-7816, 2020.

[30] Z. Zhong, J. Li, Z. Luo, and M. Chapman, "Spectral–spatial residual network for hyperspectral image classification: A 3-D deep learning framework," *IEEE Trans. Geosci. Remote Sens.,* vol. 56, no. 2, pp. 847-858, 2017.

[31] W. Wang, S. Dou, Z. Jiang, and L. Sun, "A fast dense spectral–spatial convolution network framework for hyperspectral images classification," *Remote sensing,* vol. 10, no. 7, p. 1068, 2018.

[32] Z. Zhou *et al.*, "Spikformer: When spiking neural network meets transformer," *arXiv preprint arXiv:2209.15425,* 2022.

[33] Q. Yu, L. Wang, and J. Dang, "Efficient multi-spike learning with tempotron-like ltp and psd-like ltd," in *Lect. Notes Comput. Sci.*, 2018: Springer, pp. 545-554.

[34] A. Alemi, C. Machens, S. Deneve, and J.-J. Slotine, "Learning nonlinear dynamics in efficient, balanced spiking networks using local plasticity rules," in *AAAI Conf. Artif. Intell., AAAI*, 2018, vol. 32, no. 1.

[35] W.-M. Kang *et al.*, "A spiking neural network with a global self-controller for unsupervised learning based on spike-timing-dependent plasticity using flash memory synaptic devices," in *Proc Int Jt Conf Neural Networks*, 2019: IEEE, pp. 1-7.

[36] T. Bu, J. Ding, Z. Yu, and T. Huang, "Optimized potential initialization for low-latency spiking neural networks," in *AAAI Conf. Artif. Intell., AAAI*, 2022, vol. 36, no. 1, pp. 11-20.

[37] Q. Yu, C. Ma, S. Song, G. Zhang, J. Dang, and K. C. Tan, "Constructing accurate and efficient deep spiking neural networks with double-threshold and augmented schemes," *IEEE Transactions on Neural Networks and Learning Systems,* vol. 33, no. 4, pp. 1714-1726, 2021.

[38] Q. Zhan, G. Liu, X. Xie, M. Zhang, and G. Sun, "Bio-inspired Active Learning method in spiking neural network," *Knowl Based Syst.,* vol. 261, p. 110193, 2023.

[39] L. Zhu, X. Wang, Y. Chang, J. Li, T. Huang, and Y. Tian, "Event-based video reconstruction via potential-assisted spiking neural network," in *Proc IEEE Comput Soc Conf Comput Vision Pattern Recognit*, 2022, pp. 3594-3604.

[40] Y. Wu, L. Deng, G. Li, J. Zhu, and L. Shi, "Spatio-temporal backpropagation for training high-performance spiking neural networks," *Frontiers in neuroscience,* vol. 12, p. 331, 2018.

[41] Y. Wu, L. Deng, G. Li, J. Zhu, Y. Xie, and L. Shi, "Direct training for spiking neural networks: Faster, larger, better," in *AAAI Conf. Artif. Intell., AAAI*, 2019, vol. 33, no. 01, pp. 1311-1318.

[42] F. Zenke and T. P. Vogels, "The remarkable robustness of surrogate gradient learning for instilling complex function in spiking neural networks," *Neural Comput Appl,* vol. 33, no. 4, pp. 899-925, 2021.

[43] X. Cheng, Y. Hao, J. Xu, and B. Xu, "LISNN: Improving spiking neural networks with lateral interactions for robust object recognition," in *IJCAI*, 2020: Yokohama, pp. 1519-1525.

[44] Y. Liu, K. Cao, R. Wang, M. Tian, and Y. Xie, "Hyperspectral image classification of brain-inspired spiking neural network based on attention mechanism," *IEEE Geosci. Remote Sens. Lett.,* vol. 19, pp. 1-5, 2022.

[45] Y. Liu, K. Cao, R. Li, H. Zhang, and L. Zhou, "Hyperspectral Image Classification of Brain-Inspired Spiking Neural Network Based on Approximate Derivative Algorithm," *IEEE Trans. Geosci. Remote Sens.,* vol. 60, pp. 1-16, 2022.

[46] K. Roy, A. Jaiswal, and P. Panda, "Towards spike-based machine intelligence with neuromorphic computing," *Nature,* vol. 575, no. 7784, pp. 607-617, 2019.

[47] J. M. Brader, W. Senn, and S. Fusi, "Learning real-world stimuli in a neural network with spike-driven synaptic dynamics," *Neural Comput Appl,* vol. 19, no. 11, pp. 2881-2912, 2007.

[48] R. Luke and D. McAlpine, "A spiking neural network approach to auditory source lateralisation," in *ICASSP IEEE Int Conf Acoust Speech Signal Process Proc*, 2019: IEEE, pp. 1488-1492.

[49] E. M. Izhikevich, "Simple model of spiking neurons," *IEEE Trans Neural Netw Learn Syst,* vol. 14, no. 6, pp. 1569-1572, 2003.

[50] A. L. Hodgkin and A. F. Huxley, "A quantitative description of membrane current and its application to conduction and excitation in nerve," *J. Physiol.,* vol. 117, no. 4, p. 500, 1952.

[51] L. F. Abbott, "Theoretical neuroscience rising," *Neuron,* vol. 60, no. 3, pp. 489-495, 2008.